\documentclass[aps,prb,reprint,preprintnumbers,amsmath,amssymb,floatfix]{revtex4-2}

\usepackage{longtable}
\usepackage{morefloats}
\usepackage{braket}
\usepackage{color}
\usepackage{epsfig,graphicx,amsfonts,amsbsy}
\usepackage{amsmath,amsfonts,amsthm,amssymb}
\usepackage{appendix}
\usepackage{bbm}
\usepackage{makeidx}
\usepackage{url}
\usepackage{verbatim}
\usepackage[bookmarksnumbered,pdfpagelabels=true,plainpages=false,colorlinks=true,linkcolor=blue,citecolor=blue,urlcolor=blue]{hyperref}
\usepackage{mathtools}
\usepackage[rightcaption]{sidecap}
\usepackage{array}
\usepackage{booktabs}
\usepackage{multirow}
\usepackage{floatrow}
\usepackage{tabularx}
\usepackage{bm}
\usepackage{gensymb}
\usepackage{CJK}
\makeatletter
\newsavebox{\@brx}
\newcommand{\llangle}[1][\scriptstyle]{\savebox{\@brx}{\(\m@th{#1\langle}\)}%
  \mathopen{\copy\@brx\kern-0.5\wd\@brx\usebox{\@brx}}}
\newcommand{\rrangle}[1][\scriptstyle]{\savebox{\@brx}{\(\m@th{#1\rangle}\)}%
  \mathclose{\copy\@brx\kern-0.5\wd\@brx\usebox{\@brx}}}
\makeatother

\begin{document}
\title{High bosonic Bott index and transport of multi-band topological magnons}
\author{Kai-Tao Huang}
\author{X.S.Wang}
\email{justicewxs@hnu.edu.cn}
\affiliation{School of Physics and Electronics, Hunan University, Changsha 410082, China}

\begin{abstract}
Magnons are bosonic quasiparticles in magnetically ordered systems. Bosonic Bott index has been affirmed as a real-space topological invariant for a two-band ferromagnetic model. In this work, we theoretically investigate the topology and transport of magnons in a multi-band bosonic Kagome ferromagnetic model. We demonstrate the validity of the bosonic Bott indices of values larger than 1 in multi-band magnonic systems by showing the agreement with Chern numbers in the clean limit and the bulk-boundary correspondence during the topological phase transition. For the high Bott index phase, the disorder-induced topological phase transition occurs in a multi-step manner. Using a generalized Landauer-B\"{u}ttiker formalism, we reveal how the magnon transport depends on Gilbert damping and disorder under coherent excitation or temperature difference.
The results further justify the bosonic Bott index as a robust real-space topological invariant for multi-band magnonic systems and provide insights into the transport of topological magnons.

\end{abstract}
\maketitle

\section{Introduction}
Magnons, quantized spin-wave excitations first proposed by F. Bloch \cite{Bloch1930}, represent collective quasiparticles in magnetically ordered systems. Due to the low energy dissipation and long coherence lengths \cite{Cornelissen2015,Lebrun2018,Liu2018,Pirro2021,Hu2024}, magnons have emerged as promising candidates for information carriers in next-generation spintronic devices \cite{Khitun_2010,Sorin2020,Vogt2014,Jamali2013,Klingler2015}. These advances have promoted the development of \textit{magnonics}, focusing on excitation \cite{Xiao2013,Zivieri2018,vcharov2024}, propagation \cite{Zhang2014,Li2024,Moon2013,zeng2025}, manipulation \cite{Eguchi2022,Grachev2023,Toedt2021}, and detection \cite{Borst2023,Balinskiy2021,Lucassen2023} of magnons, with the aim of designing and implementing of magnon-based devices.

In the last two decades, topological classification of condensed matter has attracted much attention, partially because the non-trivial band topology correlates with exotic transport properties, such as robust edge states \cite{TI1,TI2} and anomalous quantum Hall effect \cite{AQH}. Such topological classification, which was studied in electronic systems at the beginning, has now been successfully extended to bosonic excitations, including photons \cite{Khanikaev2013,Yang2019,Cardano2017}, phonons\cite{Yang2023,Yu2018}, and magnons \cite{Lifa2013,Mook2014,Chisnell2015,Owerre_2016,Wang2017,Owerre2018,Wang2018,Wei_2022,Zhang2023,Zhuo2024}. In topological magnonic systems, there are robust magnon edge states that may be useful in magnonic devices \cite{Wang2018}. For example, the synergy between the Dzyaloshinskii-Moriya interaction (DMI) \cite{DZYALOSHINSKY1958241,Moriya1960} and the Heisenberg exchange interaction induce non-trivial magnon band topology in Kagome \cite{Lifa2013,Mook2014}, honeycomb \cite{Owerre_2016}, triangular Kagome \cite{Zhang2023}, and Lieb lattices \cite{Zhuo2024}.  However, in these studies, the single-particle magnon Hamiltonians can be mapped to known fermionic topological models such as the Haldane model \cite{Haldane1988}. Indeed, the bosonic nature of magnons is embodied not only by the multi-particle Bose-Einstein statistics, but also by the single-particle bosonic commutation relations. Once the single-particle Hamiltonian $H$ contains terms like $a_1a_2$ and $a_1^\dagger a_2^\dagger$ ($a_1$, $a_2$ and $a_1^\dagger$, $a_2^\dagger$ are annihilation and creation operators, respectively), the Bogoliubov transformation gives a non-trivial metric $\eta$ leading to a pseudo-Hermitian eigenvalue problem $\eta H \Psi =\Psi \eta E$ \cite{White1965,Wang2021}, that cannot be mapped to any fermionic systems. Therefore, the topology of magnonic systems with discrete translational symmetry has to be characterized by the bosonic version of Chern number \cite{Shindou2013} in the reciprocal space. Meanwhile, systems lacking translation symmetry are also ubiquitous, such as disordered systems, quasicrystals, and amorphous systems. Developing tools for real-space topological classification is essential for such systems. Bosonic Bott index has been affirmed to be a real-space topological invariant in a two-band model with onset disorder \cite{Wang2020}. Non-trivial Bott index of values $\pm1$ was shown to be able to characterize the magnon band topology. However, although there are examples of high Chern number magnons \cite{SkyrLattice2016,Yao2023, Zou2025}, whether the bosonic Bott index still works for multi-band systems has not been demonstrated yet. 

On the other hand, magnon transport is the key feature that directly relates to device applications. There are a number of studies concerning the interplay of disorder and topology in electric systems \cite{TAI1,TAI2,Titum2015,Zheng2024,Chowdhury,Oliveira2024}, but in magnonic systems, the physics will be distinct from the electronic counterpart, because of not only the bosonic statistics, but also the non-conservation of magnon numbers. The loss of magnon number due to various inelastic processes can be taken into account by the Gilbert damping\cite{Gilbert2004}. How topology, disorder, and damping influence magnon transport is still underexplored.

\begin{figure}[!th]
		\centering
		\includegraphics[width=\textwidth]{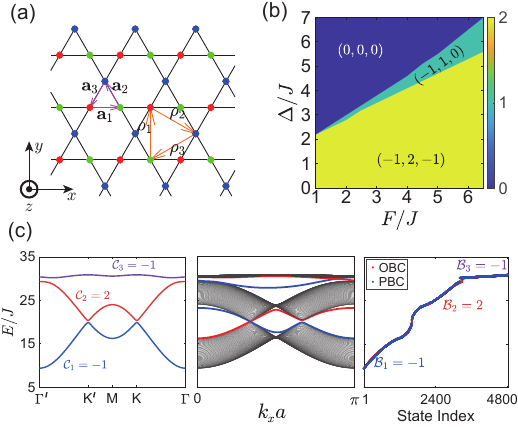}
		\caption{(a) Schematic diagram of Kagome lattice. The red,green and blue atoms represent the A,B and C sublattices in a unit cell. $\mathbf{a}_i$ and $\bm{\rho}_i$ denote the NN and NNN vectors, respectively. (b) Topological phase diagram in $F-\Delta$ parameter space. (c) Magnon spectrum for infinite sample, strip sample, and $40\times 40$ finite sample (from left to right, respectively). The parameters are  \textcolor{black}{$\Delta=J^{\prime}=0$}, $K=20J$ and $F=5J$. $\mathcal{C}_{1,2,3}$ denote Chern numbers and $\mathcal{B}_{1,2,3}$ denote Bott indices for the corresponding bands.}
		\label{fig1}
\end{figure}

In this paper, we take a step further to affirm the bosonic Bott index of absolute values larger than 1 for multi-band cases. We consider the magnons in a three-band Kagome ferromagnet \cite{Chisnell2015,KM1,Ye2018,Li2021} in the presence of pseudodipolar interaction \cite{Moriya1960,Jackeli2009} or hard-axis anisotropy along the bands, either of which results in non-trivial metric $\eta$. In the clean limit, the Bott index is consistent with the Chern number, whose maximal values can be 4. In the presence of onset disorder, the non-trivial-to-trivial topological transition accompanies the localization of edge states, showing the bulk-boundary correspondence. For high Bott index bands, the disorder-induced topological transition occurs in a multi-step manner. At last, using a generalized Landauer-B\"{u}ttiker formula \cite{zhangtian2023}, we study the magnon transport in the presence of disorder and Gilbert damping, for both coherent pumping and incoherent thermal excitation. While the increasing disorder strength reduces the magnon current, the damping affects the magnon current in different ways for the coherent pumping in the gap, coherent pumping in the bulk band, and thermal excitation. Our results further affirm the bosonic Bott index as a real-space topological index, and provide more physical insights for the practical application of topological magnonics.

\section{Bosonic Bott Index for Multiband Case}
We consider a ferromagnet with localized spins on a two-dimensional Kagome lattice as shown in Fig. \ref{fig1}(a). The total spin Hamiltonian is given by
\begin{multline}
  \mathcal{H}=\frac{1}{S}\bigg[-J\sum_{ \langle i,j \rangle} \mathbf{S}_i \cdot \mathbf{S}_j -\sum_i \frac{K_i}{2}(S^z_i)^2 \\
  \textcolor{black}{-J^{\prime}\sum_{\langle \langle i,j\rangle\rangle} \mathbf{S}_i \cdot \mathbf{S}_j}-F \sum_{\langle i,j \rangle}( \mathbf{S}_i \cdot \mathbf{e}_{ij})(\mathbf{S}_j \cdot \mathbf{e}_{ij})\bigg],
\label{total_H}  
\end{multline}
where $\mathbf{S}_i$ is the local spin at site $i$ with total spin $S$. The first term represents the nearest-neighbor (NN) ferromagnetic Heisenberg exchange interaction $J>0$.  \textcolor{black}{ The second term represents the next nearest-neighbor (NNN) ferromagnetic exchange interaction $J^{\prime}$. In this section, we set $J^\prime=0$.} The third term represents an easy-axis anisotropy along $z$ direction $(K_i>0)$ that can be different at each site. The last term is the NN pseudodipolar interaction which is an anisotropic superexchange resulting from spin-orbit coupling \cite{Moriya1960,Jackeli2009} with $F$ the coupling constant. $\mathbf{e}_{ij} $ is the unit vector connecting lattice sites $i$ and $j$ along one of the NN lattice vectors $\mathbf{a_{1,2,3}}$, corresponding to the angles to the $x$ direction $\theta_1=0$, $\theta_2=2\pi/3$ and $\theta_3=\pi/3$.

Applying the Holstein-Primakoff (HP) transformation \cite{HPtransformation} and keeping the linear terms only, the real-space single-particle magnon Hamiltonian is
\begin{multline}
\mathcal{H}= -\left(J+\frac{F}{2}\right)\sum_{\langle i,j\rangle}s_i^\dagger s_j - \textcolor{black}{J^{\prime}\sum_{\langle\langle i,j \rangle \rangle}s_i^\dagger s_j}
  \\-\frac{F}{2} \sum_{\langle i,j\rangle} \left(e^{-2i\theta_{ij}}s_i s_j +\mathrm{h.c.} \right) 
+\sum_i  \textcolor{black}{M_i} s_i^\dagger s_i \label{Full hamiltonian}
\end{multline}
where $s_i$ ($s^\dagger_i$) is the annihilation (creation) operator on site $i$ (can be one of A, B, or C), and $\theta_{ij}$ is one of $\theta_{1,2,3}$ denoting the angle of the bond between $i$ and $j$.  \textcolor{black}{$M_i=ZJ+Z^{\prime}J^{\prime}+K_i$ where $Z$ is the NN coordination number and $Z^{\prime}$ is the NNN coordination number ($Z=Z^{\prime}=4$ in the bulk and may be less than 4 at the edge and corner).} Clearly, the pseudodipolar interaction gives $s_i s_j$ and $s_i^\dagger s_j^\dagger$ terms, leading to a non-trivial bosonic metric.  For a finite sample of $N$ unit cells, if the operators are arranged as $x=\left(\ldots,s_i,s_i^\dagger,\ldots\right)^T$, the Hamiltonian can be written as $\mathcal{H}=\frac{1}{2}x^\dagger H x$. To diagonalize the $6N\times 6N$ matrix $H$, Bogoliubov transformation is used to obtain the eigenvalue problem, $\eta_r H\Psi=\Psi \eta E_r$, where $\eta_r=\mathbbm{1}_{3N} \otimes \sigma_z$ is the bosonic metric. $\Psi$ is the transformation matrix, and $E_r$ is the real-space energy spectrum. 

The Fourier transformation gives the $k$-space eigenvalue problem in a clean and infinite system $\eta H_\mathbf{k} \Psi_\mathbf{k} =\Psi_\mathbf{k}\eta E_\mathbf{k}$, where $H_\mathbf{k}$ is a $6\times 6$ matrix (See Appendix \ref{mat_detail}), and the corresponding bosonic metric is $\eta=\mathbbm{1}_3\otimes \sigma_z$. The energy spectra of infinite system, 80-wide stripe, and $40\times 40$ finite sample are plotted and compared in Fig. \ref{fig1}(c)  \textcolor{black}{(in this section, parameters $K=20J$, $F=5J$, $\Delta=0$, $J^{\prime}=0$ are used, unless otherwise specified).} Since the Bogoliubov diagonalization gives pairs of ``particle bands" and ``hole bands" which are actually equivalent, we only show the three positive bands. Three bands separated by two gaps, denoted by band 1, band 2, and band 3 from low energy to high energy, can be identified. In the stripe sample, there are a pair of edge states inside each gap. In the finite sample, the periodic boundary condition (PBC) gives finite gaps, while edge states in the gaps are present under open boundary condition (OBC).  \textcolor{black}{ The red (blue) curve in the middle panel of Fig.\ref{fig1}(c) denotes the topological edge states contributed by the lower (upper) edge of sample. }The edge states in the lower (upper) gap propagate counterclockwise (clockwise). 

The $k$-space Chern numbers and the real-space Bott indices of the three gapped bands can be calculated using the formulae given in previous studies \cite{Shindou2013,Wang2020,Wang2021}
 \textcolor{black}{
\begin{equation}
\mathcal{C}_n=\frac{i}{2\pi}  \iint_{B.Z.} \Omega_{nz}(\mathbf{k}) \mathrm{d}\mathbf{k}, \label{Chern number}
\end{equation}
\begin{equation}
\mathcal{B}_n=\frac{1}{2 \pi}\mathrm{Im} \{ \mathrm{Tr} \left[ \mathrm{log} \left( V U V^\dagger U^\dagger \right) \right] \},
\label{Bott index}
\end{equation}
where the subscript $n$ denotes the $n$st band and more details can be found in Appendix \ref{topo_inv}.} Calculation shows that the Chern numbers and Bott indices are consistent with each other, namely $-1$, $2$, and $-1$ for the three bands 1, 2, and 3, respectively. In the following, we use $(\mathcal{B}_1,\mathcal{B}_2,\mathcal{B}_3)$ to notate the topology of magnon bands. The existence of counter-propagating edge states in the two gaps resembles the bulk-boundary correspondence. We then break the equivalence of the A, B, and C sublattices by introducing a staggered anisotropy. To avoid complication, we only consider the case that $K_\mathrm{A}=K$, $K_\mathrm{B}=K+\Delta$, and $K_\mathrm{C}=K-\Delta$. Figure \ref{fig1}(b) shows the topological phase diagram in $F-\Delta$ parameter space. By increasing $\Delta$, the Bott indices can be tuned to $(-1,1,0)$, and $(0,0,0)$ successively. For the $(-1,1,0)$ phase, robust edge states exist only in the lower gap, while for the $(0,0,0)$ phase there is no edge state (not shown). The bosonic Bott index of largest value 2 and the bulk boundary correspondence in clean systems are affirmed.

\begin{figure}[!th]
		\centering
		\includegraphics[width=\textwidth]{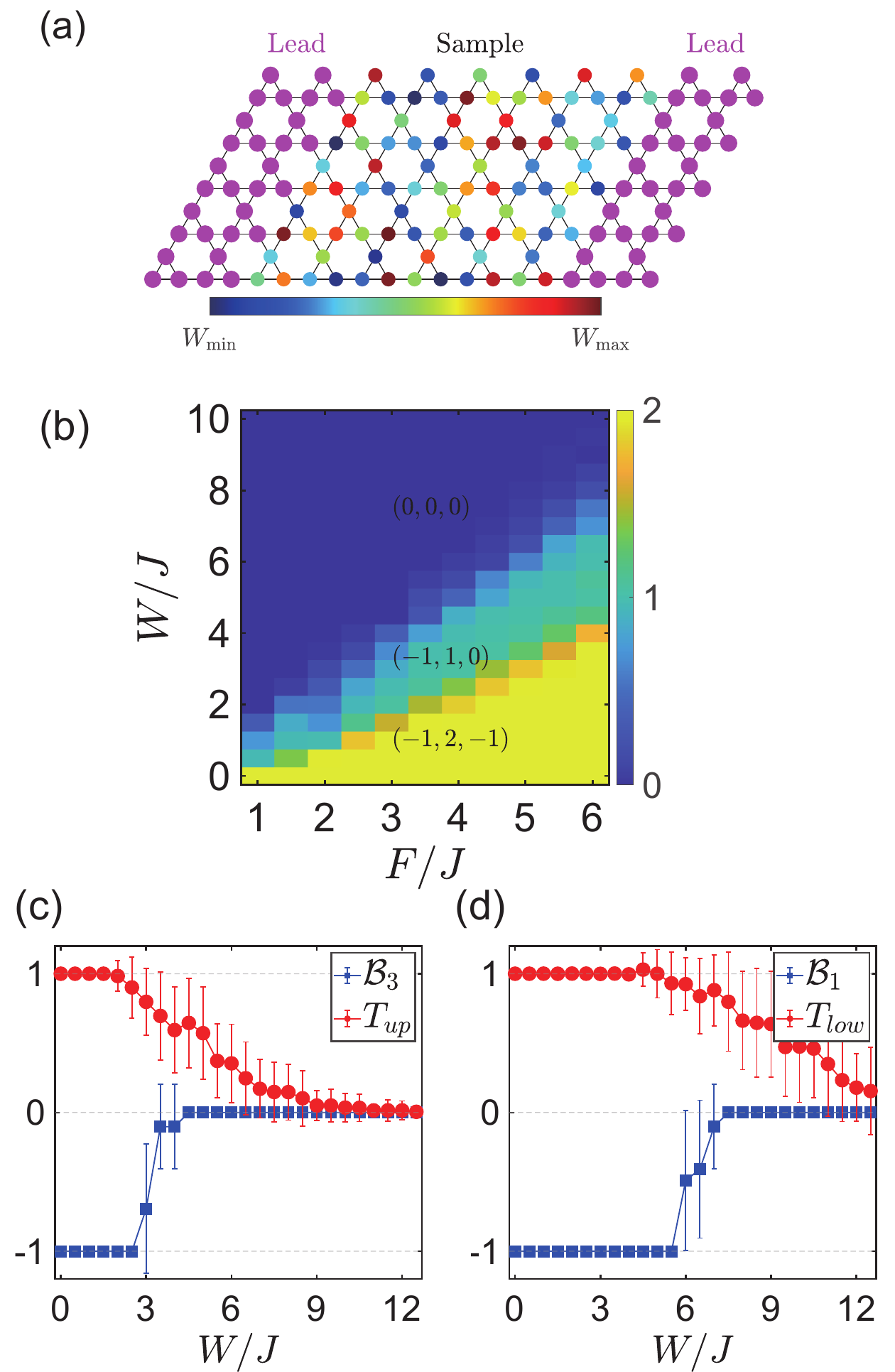}
		\caption{(a) Schematic diagram of the lead-sample-lead device for magnon transport calculation. The colors of the circles illustrate the anisotropy at the spin sites. (b) Topological phase with respect to pseudodipolar interaction strength $F$ and disorder strength $W$. (c)(d) Magnon transmission $T$ calculated at the center of the (c) upper gap and (d) lower gap, compared with the Bott index of (c) the third band and (d) the first band. The energy values for transmission function calculation are $20.11J$ for lower gap and $29.65J$ for upper gap, respectively.  \textcolor{black}{Here $F=5J$, $K=20J$, and $\Delta=J^\prime=0$. 50 samples are averaged.} $\mathcal{C}_{1,2,3}$ and $\mathcal{B}_{1,2,3}$ denote the Chern numbers and Bott indices of the corresponding bands.}
		\label{fig2}
\end{figure}

Disorder is ubiquitous in reality, and the interplay between disorder and topology is a meaningful issue \cite{TAI1,TAI2,Wang2020}. The presence of disorder breaks the discrete translational symmetry of the lattice so that the $k$-space Chern number is not valid, and the Bott index becomes a more suitable topological index. We consider onset disorder by adding a random $\Delta K$ to the anisotropy constant $K$ of each site. $\Delta K$ follows a uniform distribution in the range $[-W,W]$. The topological phase diagram for $K=20J$ in $F-W$ parameter space is plotted in Fig. \ref{fig2}(b). PBC is applied for a $20\times 20$ sample while 50 random sets are averaged for the Bott index calculation. By increasing the strength of disorder, topological phase transition occurs from $(-1,2,-1)$ to $(-1,1,0)$, and then $(0,0,0)$. To further demonstrate the bulk-boundary correspondence, we consider a typical lead-sample-lead set-up for transport calculation, as shown in Fig. \ref{fig2}(a). The disordered anisotropy exists in the sample at the middle, while the left and right leads are clean with the same $K=20J$ and $F=5J$. Using the standard nonequilibrium Green's function method \cite{ryndyk2016theory}, the total magnon transmission $T$ across the sample from the left lead to the right lead at a certain energy $E$ can be calculated, while its number equals the total number of extended states in ballistic transport. For $\Delta=0$ which is in the $(-1,2,-1)$ phase, the magnon transmission at $E=29.65J$ which is inside the upper gap is compared with the Bott index of band 3 (average of 50 samples) in Fig. \ref{fig2}(c). When the Bott index is nontrivial with value -1, there is one rightward extended edge state, giving total transmission 1. When the disorder is large enough and the Bott index becomes 0, the edge state is also localized, leading to a transmission decay to 0. For $E=20.11J$ in the lower gap, similar nontrivial-to-trivial transition occurs at stronger disorder strength because of the large gap, as shown in Fig. \ref{fig2}(d). However, the extended-localized transition of magnon edge states is smoother that the topological transition labelled by the Bott index. The reason may be attributed to the finite-size effect. If the sample is not long enough, even if all the states are localized, there will still be finite transmission, because the localization length of some states may be comparable to the sample length. On the other hand, if the sample is not wide enough, the edge states at two edges will overlap each other, providing a back-scattering channel, so that the transmission becomes less than 1. Limited by our computational capacity, we show the results for 20 (long)$\times$80 (wide) samples. If the system is large enough, the transition will be sharper (see the Appendix \ref{size_eff} for details).

\begin{figure}[!th]
		\centering
		\includegraphics[width=\textwidth]{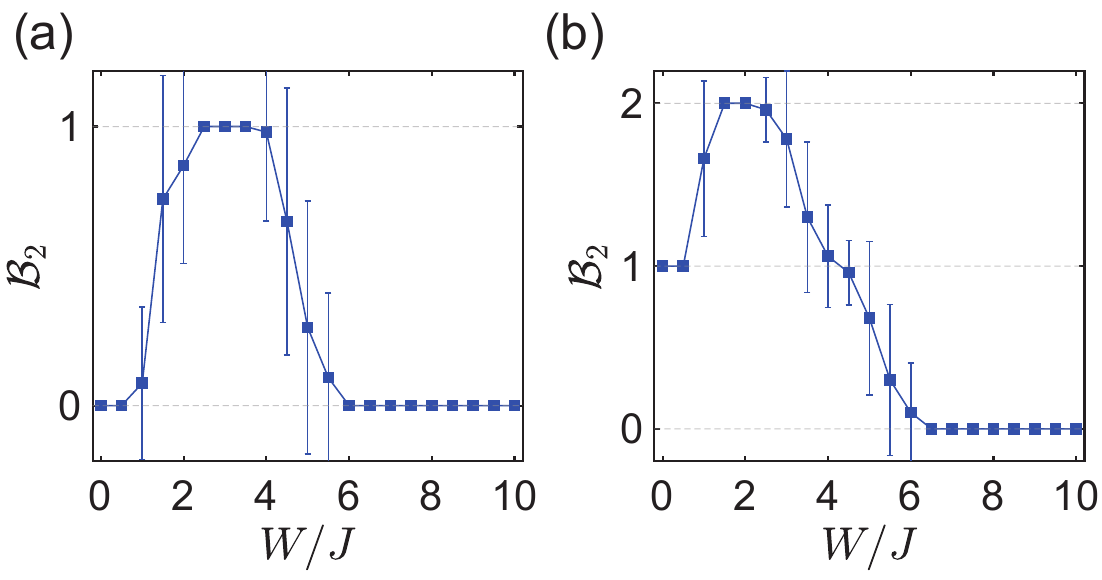}
		\caption{The Bott index of the second band $\mathcal{B}_2$ for different disorder strength for (a) $\Delta=4.7J$ [$(0,0,0)$ phase when $W=0$], and (b) $\Delta=5.5J$ [$(-1,1,0)$ phase when $W=0$]. Sample size is $30 \times 30$ and 50 samples are averaged and other parameters are \textcolor{black}{ $F=5J$, $K=20J$, $J^{\prime}=0J$.}}
		\label{fig3}
\end{figure}

For $\Delta=5.5J$, $F=5J$ case which belongs to the trivial $(0,0,0)$ phase  near the phase boundary in Fig. \ref{fig1}(b), the increasing disorder drives a topological transition to the nontrivial phase $(-1,1,0)$ with a plateau of $\mathcal{B}_2=1$. For $\Delta=4.7J$, $F=5J$ which belongs to the $(-1,1,0)$ phase near the phase boundary in the absence of disorder, Fig. \ref{fig3} plots $\mathcal{B}_2$, the Bott index of band 2 (the middle band), against the disorder strength $W$. The Bott index starts from 1 and transits to 2. For stronger disorder, the Bott index transits back to 1 and then 0. Similar disorder-induced trivial-to-nontrivial topological transition has been observed in the well-known topological Anderson insulator \cite{TAI1,TAI2} and two-band magnon models \cite{Wang2020}. The disorder in the onset energy (i.e. the anisotropy here) effectively reduces the discrepancies between different sublattices in the form of a mean-field self energy \cite{TAI2,Wang2020,Han2023}, so that the topological transition occurs.

\section{Transport of Topological Magnons}
\begin{figure}[!th]
		\centering
		\includegraphics[width=\textwidth]{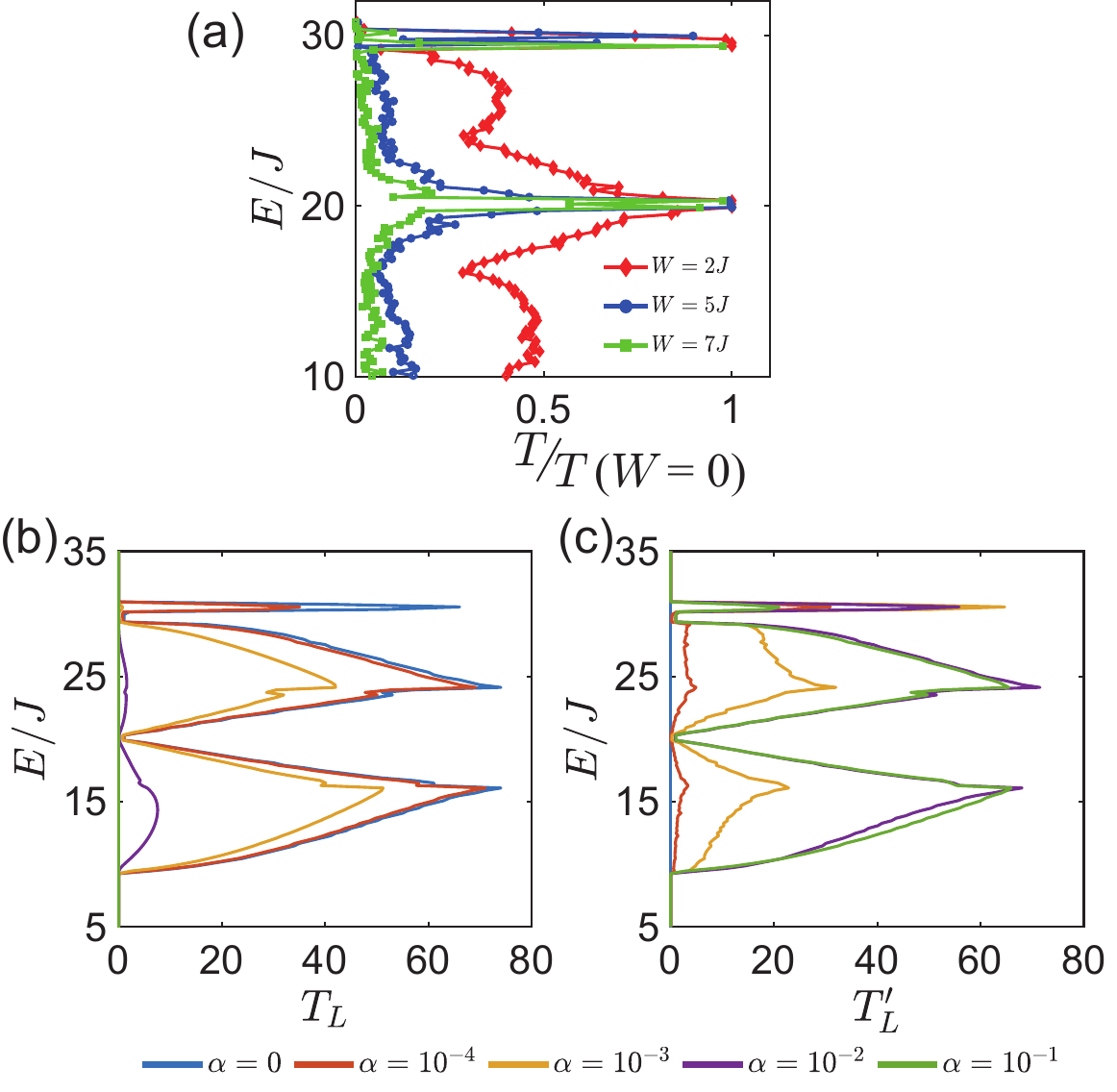}
		\caption{(a) The magnon transmission $T$ across the device divided by the value of clean case $T(W=0)$ for different disorder strength with $\Delta=0$. (b)(c) The transmission functions for $W=0$ and different damping $\alpha$. (b) The transmission from left lead to right lead $T_l$. (c) The transmission from left lead to central sample $T_l^{\prime}$.  \textcolor{black}{Here $F=5J$, $K=20J$, and $\Delta=J^\prime=0$}. }
		\label{fig4}
\end{figure}

To utilize the topological magnons, it is necessary to study the transport properties of magnons. Unlike fermionic electrons that only states near the Fermi level contribute to the current, the magnons follow the bosonic distribution so that all the low-energy states contribute to the current. Thus, although we have shown that the bulk-boundary correspondence also applies for the topological magnons for high-Bott-index case, the transport properties of magnons are extremely different from the electronic counterpart. The magnon current $j_l$ from the left lead to the sample and $j_r$ from the right lead to the sample are expressed as \cite{zhangtian2023},
\begin{equation}
\begin{aligned}
& j_{l(r)}=\int \frac{dE}{2\pi} \left[f_{l(r)} \left({E} \right)-f_{r(l)} \left({E} \right) \right]T_{L(r)}(E) \\
& +\int \frac{dE}{2\pi}\left[f_{l(r)} \left({E})-f_{c}({E} \right) \right] T^{\prime}_{l(r)}(E), \label{LBF}
\end{aligned}
\end{equation}
where the subscripts $l$, $r$, $c$ correspond to the left lead, the right lead, and the central sample, respectively. $T_{l(r)}(E)=\mathrm{Tr}[\Gamma_{l(r)} G^R \Gamma_{r(l)} G^A]$ is the transmission function \cite{zhangtian2023, ryndyk2015book} from the left (right) lead to the right (left) lead, and $T^{\prime}_{l(r)}(E)=\mathrm{Tr}[\Gamma_{l(r)} G^R \Gamma_{c} G^A]$ is the transmission function from the left (right) lead to the central sample (more details can be found in Appendix \ref{NEGF}).  $f$ is the magnon distribution function in each region. The integration is over the magnon energy $E$. For damping $\alpha=0$, the current across the whole device is conserved, so $T_l=T_r=T$ and $T^\prime_{l(r)}=0$. We plot the transmission $T$ divided by the value of the clean sample $T(W=0)$ for different $E$ in Fig. \ref{fig4}(a), for $W=2J, 5J$ and $7J$. It is clear that, the ratio is closed to 1 in the gaps, while in the bulk bands the ratio decays to 0 for larger disorder, clearly demonstrating the robustness of the edge states. To show the effect of non-zero damping $\alpha\neq0$, the $T_l$ and $T^\prime_l$ for the clean sample are plotted in Fig. \ref{fig4}(b) and (c), respectively. The bulk bands appear as peaks since the transmission is related to the density of states. Due to the finite damping in the central sample, it acts as a sink for the magnons and a portion of magnon currents are absorbed. So, for increasing damping, $T_l$ decays while $T_l^\prime$ increases rapidly. 

To generate a magnon current across the device, nonequilibrium magnon distribution has to be induced. We consider two types of excitation for magnon currents, coherent external excitation and noncoherent temperature difference. For an external oscillating field excitation inside the left lead, the excited magnon number follows a Lorentzian function \cite{FMR1, FMR2,RezendeBook}, written as
\begin{equation}
    L(E,E_0,\Gamma) =\frac{\Lambda J}{\pi} \frac{\Gamma}{(E-E_0)^2+\Gamma^2},
\end{equation}
where $E_0/\hbar$ is the frequency of external excitation. $\Gamma=\Gamma_0+\alpha E_0$ is the linewidth, where $\Gamma_0$ is the extrinsic linewidth and $\alpha E_0$ is the damping-induced linewidth. Here, we use $\Gamma_0=0.1J$ as an example, which can be determined experimentally in reality \cite{Chisnell2015}. The distribution functions are $f_l(E)=N_B(E/k_Bt_l)+ \textcolor{black}{\Lambda} L(E)$, $f_{ \textcolor{black}{c}}(E)=N_B(E/k_Bt_{c})$, and $f_r(E)=N_B(E/k_Bt_r)$ in the left lead ($l$), central sample ($ \textcolor{black}{c}$), and right lead ($r$), respectively.  $t_{l, \textcolor{black}{c},r}$ is the temperature, and  \textcolor{black}{$\Lambda$} is an amplitude proportional to the input power of the excitation. $N_B$ is the Bose-Einstein distribution function. The temperature of the central sample is set to $k_Bt_{ \textcolor{black}{c}}=43J$, which corresponds to room temperature for a reasonable value of $J=0.6$ meV \cite{Exp2021}. 

\begin{figure}[!th]
		\centering
		\includegraphics[width=\textwidth]{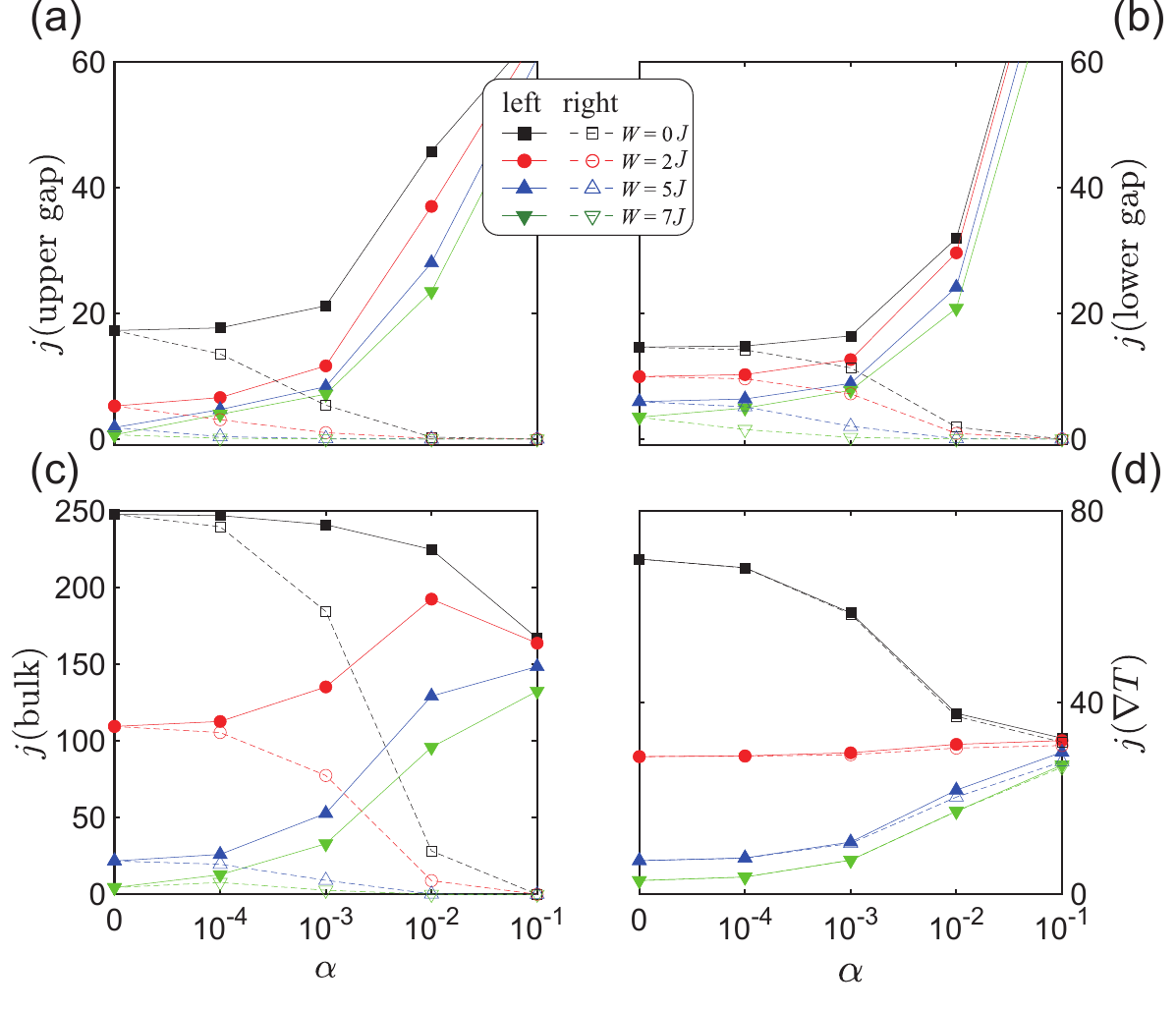}
		\caption{Magnon currents at the two interfaces for different damping and disorder strength for different excitations. (a)(b)(c) External frequency in the left lead of (a) $E=29.65J$ (in the upper gap), (b) $E=20.11J$ (in the lower gap), and (c) $E=14.5J$ (in the bulk band). (d) Temperature difference $t_l=1.2t_{ \textcolor{black}{c}}$ and $t_R=0.8t_{ \textcolor{black}{c}}$.  \textcolor{black}{Here $F=5J$, $K=20J$, and $\Delta=J^\prime=0$.} }
		\label{fig5}
\end{figure}

The magnon currents at the left interface $j_l$ (solid symbols) and right interface $j_r$ (hollow symbols) for different damping and disorder strength under different excitations are shown in Fig. \ref{fig5}. Figure \ref{fig5}(a) and (b) shows the results under coherent external excitations of frequency $E_0=29.65J$ and $E_0=20.11J$, which are in the upper gap and the lower gap to excite the robust edge states. $\Lambda$ is set to be 22.5,  \textcolor{black}{corresponding to a reasonable pumping rate (see Appendix \ref{cur_detail} for detailed discussion).} For $\alpha=0$, $j_l=j_r$, and as the damping increases, $j_l$ increases and $j_r$ decays rapidly for both gaps. That is because for larger $\alpha$, the linewidth of excited magnons is larger. When the linewidth is comparable with the gap size, a number of bulk states are excited and contribute to $j_l$. However, because of  damping, the magnon current decays in the central sample, and finally only a small proportion of magnons can penetrate to the right lead, leading to decreasing $j_r$ (see Fig. \ref{fig9} in Appendix \ref{cur_detail}). In the presence of disorder, some of the magnons are localized, so for a stronger disorder, both $j_l$ and $j_r$ decrease. For $W=0,2$ in the nontrivial phase, there are significantly finite currents into the right lead for $\alpha$ up to 0.001, which is reasonable for many magnetic materials. Figure \ref{fig5}(c) shows $j_l$ and $j_r$ under external frequency $E_0=14.5J$, which is in the lowest bulk band. The current
magnitudes are much larger than that in (a) and (b), and $j_l$ is nonmonotonic with $\alpha$ due to the competition between the linewidth broadening and the dissipative decay. Figure \ref{fig5}(d) shows noncoherent excitation by a temperature difference $t_l=1.2t_{c}$, $t_R=0.8t_{c}$, without external pumping. In this case, both bulk states and edge states are excited following the Bose-Einstein distribution. In this case, $j_l$ and $j_r$ do not differ too much, because the temperature differences at the two interfaces are the same. For $W=0$, all the states are extended, so the magnon current decreases when $\alpha$ increases due to the dissipation. But for localized states in the presence of disorder, a larger damping provides another channel via $T_{l(r)}^\prime$, so the magnon currents slightly increase with the damping. These results indicate that utilizing the robust magnon edge states is a highly nontrivial issue depending on how the magnon current is excited, which is in contrast to the edge states of topological insulators that can be selected exclusively by properly tuning the Fermi surface.

\section{Higher Bott Indices}
\begin{figure*}[!th]
		\centering
		\includegraphics[width=\textwidth]{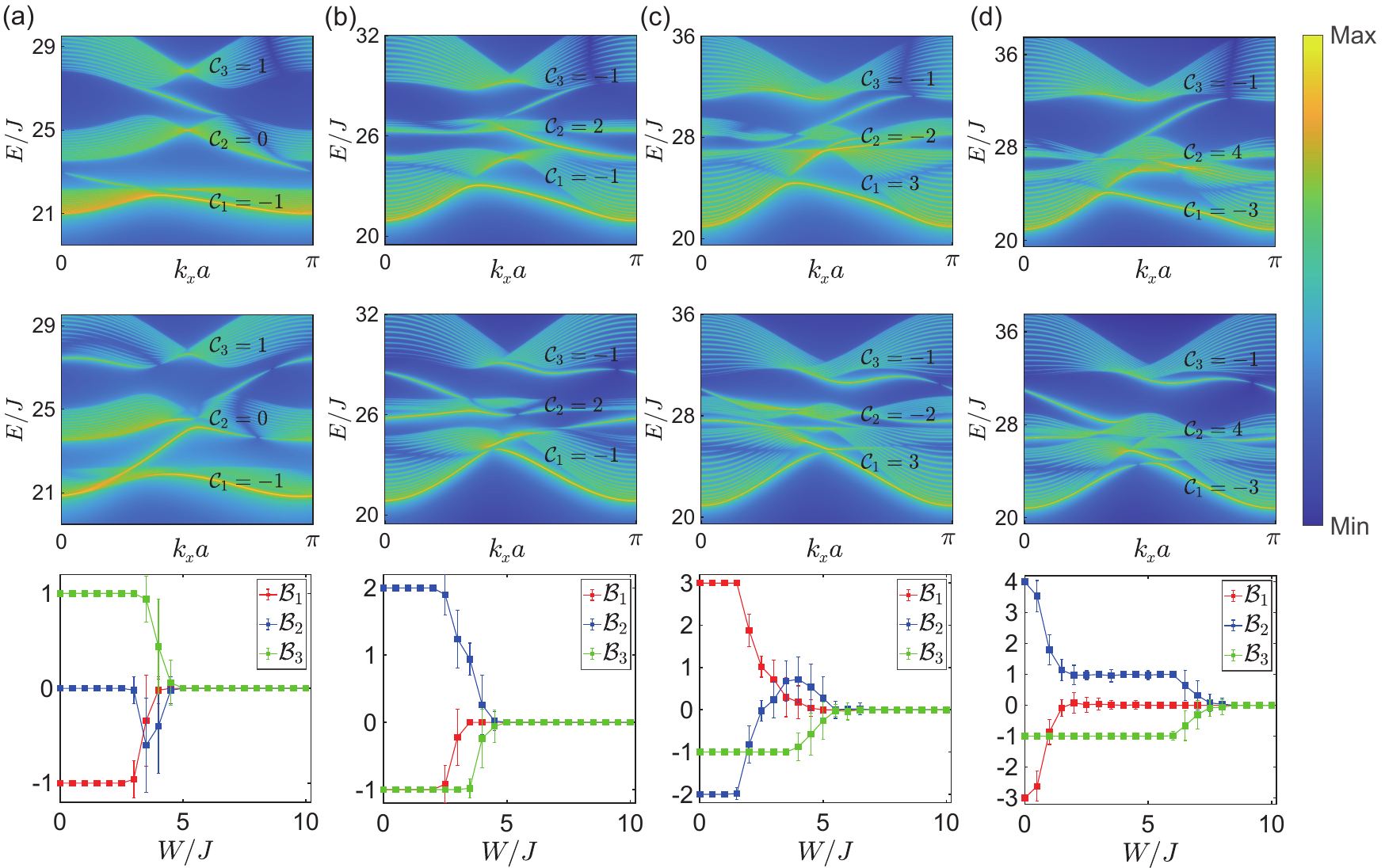}
		\caption{(a) Topological phase $(-1,0,1)$ with $D=J,J^\prime=0J$. (b) Topological phase $(-1,2,1)$ with $D=J$ and $J^\prime=0.5J$. (c) Topological phase $(3,-2,-1)$ with $D=J$ and $J^\prime=J$. (d) Topological phase $(-3,4,-1)$ with $D=1.5J$ and $J^\prime=J$. Upper panel and middle panel show the local density of states at top edge and bottom edge, respectively. The lower panel shows the Bott index of the middle band against the strength of onset disorder.  \textcolor{black}{Here $F=\Delta=0$.} }
		\label{fig6}
\end{figure*}

It is well known that the DMI $D\sum_{\langle i,j \rangle}\mathbf{d}_{ij}\cdot(\mathbf{S}_i \times \mathbf{S}_j)$  ($\mathbf{d}_{ij}$ is the $z$-projection of DMI vector\cite{PhysRevB.89.134409} and $D$ is the strength) leads to nontrivial topological magnon bands with high Chern numbers \cite{Lifa2013,Mook2014,Chisnell2015}. But in these studies, the metric $\eta$ is trivial in the magnon Hamiltonian, which means that the fermionic formula can be used directly. To highlight the bosonic nature, we consider an additional hard-axis anisotropy along the bonds at each site $i$, namely $K_h \sum_j{(\mathbf{e}_{ij} \cdot \mathbf{S}_i)^2}$ where $j$ is a nearest neighbor of $i$, meaning that the spins tend to avoid the direction along the bonds. This term is available phenomenologically \cite{Jin2012} and enters the Hamiltonian as
\begin{equation}
    \mathcal{H}_{K_h}=\frac{K_h}{2}\sum_{i,j}\begin{pmatrix}
        s_i^\dagger & s_i
        
    \end{pmatrix}\begin{pmatrix}
        1 & e^{-2i\theta_j} \\
        e^{2i\theta_j} & 1
        
    \end{pmatrix}\begin{pmatrix}
        s_i \\
        s^\dagger_i
        
    \end{pmatrix},
\end{equation}
which induces a nontrivial $\eta$. We note that other types of anisotropy, such as an in-plane uniaxial anisotropy that can be induced by strain \cite{anis2025}, may also lead to nontrivial $\eta$. In the presence of next-nearest-neighbor exchange (NNN) $\sum_{\llangle i,j\rrangle}J^\prime \mathbf{S}_i\cdot\mathbf{S}_j$, the Chern numbers (as well as the bosonic Bott indices) of the three bands can be $(-1,0,1)$, $(-1,2,-1)$, $(3,-2,-1)$ and $(-3,4,1)$. The agreement with Chern number and bulk-boundary correspondence is also verified. Figure \ref{fig6} shows the local density of states at the top edge (upper row) and bottom edge (middle row) of a stripe sample for different topological phases ($K_h=0.5J$ and $K=20J$), where the agreement between the numbers of edge states and topological invariants is affirmed. In the presence of onset disorder, topological transitions also occur in a multi-step manner. For $(-1,0,1)$ phase, there is a probability to transit to $(0,-1,1)$ for disorder around $W=4J$. For the other three phases, the system  first transits to the $(0,1,-1)$ phase. This observation indicates that the disorder-induced topological transition occurs at the lower gap first, which is different from the pseudodipolar interaction induced topology discussed in previous sections. Of course, when disorder is strong enough, the topology turns trivial for all cases.

\section{Conclusion}
To conclude, we investigate the bosonic Bott indices of magnons in a Kagome ferromagnet with nontrivial bosonic metric. We demonstrate the applicability of the bosonic Bott index with value higher than 1 for characterizing topological phases in multiband magnonic systems, showing agreement with Chern numbers in the clean limit. In the presence of onset disorder, we show the versatile disorder-induced topological transitions, and elucidate the bulk-boundary correspondence by comparing the Bott index and the transmission function. We also clarify how damping and disorders affect the magnon current for both coherent external excitation and noncoherent thermal excitation,  giving guidance to the utilization of the robust magnon edge states.

\begin{acknowledgments}
This work was supported by the National Natural Science Foundation of China (Grants No. 11804045 and 12174093), the Natural Science Foundation of Hunan Province of China (Grant No. 2025JJ60001), and the Fundamental Research Funds for the Central Universities.
\end{acknowledgments}

\appendix
\section{Calculation of Energy Spectra}\label{mat_detail}

With all the terms considered in the main text, the spin Hamiltonian Eq. \eqref{total_H} in the main text can be extended as follows,
\begin{multline}
  \mathcal{H}=\frac{1}{S}\bigg[-J\sum_{ \langle i,j \rangle} \mathbf{S}_i \cdot \mathbf{S}_j -F \sum_{\langle i,j \rangle}( \mathbf{S}_i \cdot \mathbf{e}_{ij})(\mathbf{S}_j \cdot \mathbf{e}_{ij})\\
  -J^{\prime}\sum_{\langle \langle i,j\rangle\rangle} \mathbf{S}_i \cdot \mathbf{S}_j+\sum_{ \langle i,j \rangle}\mathbf{D}\cdot(\mathbf{S}_i \times \mathbf{S}_j)\\-\sum_i \frac{K_i}{2}(S^z_i)^2+K_h \sum_{\langle i,j \rangle}{(\mathbf{e}_{ij} \cdot \mathbf{S}_i)^2}
\bigg],
\label{total_H1}    
\end{multline}
Applying HP transformation to Eq.\eqref{total_H1}  and keeping the linear terms only, the Hamiltonian can be written as

\begin{multline}
 \mathcal{H}= -\left(J+i\xi_{ij}D +\frac{F}{2}\right)\sum_{\langle i,j\rangle}s_i^\dagger s_j 
 -J^{\prime}\sum_{\langle\langle i,j \rangle \rangle}s_i^\dagger s_j \\
 -\frac{F}{2} \sum_{\langle i,j\rangle} \left(e^{-2i\theta_{ij}}s_i s_j +\mathrm{h.c.} \right)  \\
+\sum_i  \textcolor{black}{M_i} s_i^\dagger s_i+\frac{K_h}{2}\sum_{i,j}\left(e^{-2i\theta_{ij}}s_i s_i+\mathrm{h.c.}\right) 
\end{multline}
where $\xi_{ij}=\pm1$ corresponds to counterclockwise ($+$) or clockwise ($-$)  hopping from $i$ to $j$ and $D$ is the $z$-component of DMI vector. Fourier transform is performed to obtain the $k$-space Hamiltonian
\begin{equation}
H_{\mathbf{k}}=\begin{pmatrix}
 \textcolor{black}{M} & h_{13}^\dagger & f_1^- & g_1^- & f^{+}_{3} & g_3^- \\
h_{13} &  \textcolor{black}{M} & g_1^+ & f_1^+ & g_3^+ &f^{-}_{3} \\
 f^{+}_{1} & g_1^- &  \textcolor{black}{M} & h_{12}^\dagger &  f_2^- & g_2^- \\
g_1^+ &  f^{-}_{1} & h_{12} &  \textcolor{black}{M} & g_2^+ & f_2^+ \\
f_3^- & g_3^- & f^{+}_{2} &  g_2^- &  \textcolor{black}{M} & h_{23}^\dagger \\
g_3^+ & f_3^+ & g_2^+ & f^{-}_{2} & h_{23} &  \textcolor{black}{M} 
\end{pmatrix},
\label{kHamiltonian}
\end{equation}
where  \textcolor{black}{$M=4J+4J^{\prime}+K+2K_{h}$}, $f_j^{\pm}=-2(J\pm iD+F/2)\cos(\mathbf{k}\cdot \mathbf{a}_j)-2J^{\prime}\cos(\mathbf{k}\cdot \bm{\rho}_j)$, $g_j^\pm=-F e^{\pm 2i\theta_j}\cos(\mathbf{k}\cdot \mathbf{a}_j)$ and  $j=1,2,3$. $h_{jk}=K_h(e^{2i\theta_j}+e^{2i\theta_{k}})$ is the hard axis anisotropy along the bonds at each site $i$ with $j$ and $k$ being the nearest neighbor atoms of $i$. $\mathbf{a}_i$ and $\bm{\rho}_i$ are the NN and NNN vectors illustrated in Fig. 1(a) in the main text. The operators of $H_{\mathbf{k}}$ are $x=\left(s^\mathrm{A}_\mathbf{k},s_\mathbf{k}^{\mathrm{A}\dagger},s^\mathrm{B}_\mathbf{k},s_\mathbf{k}^{\mathrm{B}\dagger},s^\mathrm{C}_\mathbf{k},s_\mathbf{k}^{\mathrm{C}\dagger} \right)^T$. To diagonalize $H_{\mathbf{k}}$, Bogoliubov transformation is used to obtain the eigenvalue problem $\eta H_{\mathbf{k}}\Psi_{\mathbf{k}}=\Psi_{\mathbf{k}} \eta E_{\mathbf{k}}$, where $\eta=\mathbbm{1}_{3} \otimes \sigma_z$ is the bosonic metric. $\Psi_{\mathbf{k}}$ is the transformation matrix, and $E_{\mathbf{k}}$ is the $k$-space energy spectra. 

For $D=J^{\prime}=K_h=0$, the analytical expressions at $\mathbf{K}$ point in $k$-space can be obtained
\begin{equation}
\begin{aligned}
&  E_1=\frac{\sqrt{(6J+2K-3F)(F+6J+2K)}}{2},                    \\
&  E_2=\frac{\sqrt{(4K+18J-F)(4K+18J+3F)}}{4}-\frac{3}{4}F-\frac{1}{2}J,   \\
&  E_3=\frac{\sqrt{(4K+18J-F)(4K+18J+3F)}}{4}+\frac{3}{4}F+\frac{3}{2}J,   \\
\end{aligned}
\end{equation} 
where $E_i$ denotes the energy for $i$th band.

\section{Calculation of Topological Invariants}\label{topo_inv}

We consider the projector $P_n$ that projects an arbitrary state with momentum $\mathbf{k}$ into the $n$th band
\begin{equation}
P_n=\Psi_{\mathbf{k}} \eta \tau_n \Psi_{\mathbf{k}}^\dagger \eta,
\end{equation}
where $\tau_n$ is a matrix of the same size as $\Psi_{\mathbf{k}}$ and its $\left(n,n\right)$ is $1$ with other elements are $0$. The Chern number for $n$th bulk energy band is as the integration of its Berry curvature
\begin{equation}
\Omega_{nz}(\mathbf{k})= \mathrm{Tr} \left[ P_n \left( \frac{\partial P_n}{\partial k_x} \frac{\partial P_n}{\partial k_y}- \frac{\partial P_n}{\partial k_y} \frac{\partial P_n}{\partial k_x}\right) \right].
\end{equation}
The associated Chern number assigned to the $n$th band is given by 
Eq. \eqref{Chern number}, which is a quantized integer.

To calculate the Bott index, the real-space energy spectrum should be gapped so that all the energy eigenvalues can be divided into several groups by the gaps. Each group corresponds to an energy band in $k$ space. The PBC Hamiltonian $H_{P}$ in real space can be obtained from Eq. \eqref{Full hamiltonian} and the GEP becomes $\eta_r H_{P}\Phi=\Phi\eta_rE$. For a group of eigenvalues $\mathcal{N}$,
\begin{equation}
\mathcal{P}_{\mathcal{N}}=\Phi \eta_r \tau_{\mathcal{N}} \Phi^\dagger \eta_r
\end{equation}
projects an arbitrary state onto the set of eigenstates in $\mathcal{N}$, where $\mathcal{P}_{\mathcal{N}}=\sum_{n\in\mathcal{N}}P_n$ and $\tau_{\mathcal{N}}=\sum_{n\in\mathcal{N}}\tau_n$. One can reorder all eigenvalues in $\mathcal{N}$ after those not in $\mathcal{N}$ so that $\tau_{\mathcal{N}}$ looks like 
\begin{equation}
\tau_{\mathcal{N}}=
\begin{pmatrix}
0 & 0 \\
0 & \mathbbm{1}
\end{pmatrix},
\end{equation}
where $\mathbbm{1}$ is an identity matrix whose dimension equals to the number of eigenvalues in $\mathcal{N}$. For a set of eigenstates $E$, its bosonic Bott index for $n$th energy band is given by 
Eq. \eqref{Bott index}, and the matrices $U$ and $V$ are defined from 
\begin{equation}
\tau_{\mathcal{N}} \Phi^\dagger \eta_r e^{i 2\pi X} \Phi \eta_r\tau_{\mathcal{N}} =
\begin{pmatrix}
0 & 0 \\
0 & U
\end{pmatrix},
\end{equation}
\begin{equation}
\tau_{\mathcal{N}} \Phi^\dagger \eta_r e^{i 2\pi Y} \Phi \eta_r\tau_{\mathcal{N}} =
\begin{pmatrix}
0 & 0 \\
0 & V
\end{pmatrix},
\end{equation}
where $X=i_x /N_x$ and $Y=i_y /N_y$ are the rescaled coordinates. $i_{x,y}$ are spatial indices of the unit cell. $\mathcal{B}_n$ is always an integer when $U$ and $V$ are non-singular.

\section{Nonequilibrium Green's Function Method}\label{NEGF}
The nonequilibrium Green's function (NEGF) method is applied to study the topological magnon transport of Kagome lattice sample in real space. Its related OBC Hamiltonian matrix $H_{O}$ can also be obtained from Eq. \eqref{Full hamiltonian}. Notably, the diagonal term $M$ at the edge or corner of the sample should be modified according to coordination number. The retarded and advanced Green's function in NEGF can be written as
\begin{equation}
 \textcolor{black}{G^{R}=[E+i \epsilon^{\prime}-\eta_rH_{O}-\Sigma^{R}]^{-1},}
\end{equation}  

\begin{equation}
 \textcolor{black}{G^{A}=[E-i\epsilon^{\prime}-\eta_rH_{O}-\Sigma^{A}]^{-1}},
\end{equation}  
where $E$ is the energy of injected magnons. $\epsilon^{\prime}$ is an infinitesimal positive value and $\Sigma^{R(A)}$ is the retarded (advanced) self energy. The self energy is composed of three terms
\begin{equation}
 \textcolor{black}{\Sigma^{R}(E)=\Sigma^{R}_l+\Sigma^{R}_r+\Sigma^{R}_{c}},
\end{equation}  
where $\Sigma^R_{l(r)}$ is the electrode self energy caused by left (right) lead \cite{ryndyk2016theory} and  \textcolor{black}{ $\Sigma^R_{c}=-i\alpha E \mathbbm{1}_{6N} $ is the self energy induced by damping in the sample \cite{zhang2017,yyh2021,zhangtian2023} with $\mathbbm{1}_{6N}$ being the $6N\times6N$ identity matrix. The corresponding advanced self energy of the central sample is $\Sigma^A_{c}=i\alpha E \mathbbm{1}_{6N}$.} The lesser Green's function can be solved as
\begin{equation}
    G^<(E)=G^R(E)[\Sigma^<_l+\Sigma^<_r+\Sigma^<_{c}]G^A(E),
    \label{lesser Green function}
\end{equation}
where $\Sigma^<_s=i\Gamma_sN^B_s$  $(s=l,r, \textcolor{black}{c})$ is the lesser self energy. $\Gamma_s=i(\Sigma^R_s-\Sigma^A_s)$ is level-width function. 
The local current is given by
\begin{equation}
    j_{mn}(E)=\frac{1}{2\pi}\mathrm{Re}[G^<_{mn}(\eta_rH)_{nm}-G^<_{nm}(\eta_rH)_{mn}].
    \label{Local current}
\end{equation}
Taking the trace to sum over all the local currents and considering the distribution function, we can obtain the Landauer-like formula Eq. \eqref{LBF}, 
where  \textcolor{black}{ $T_{l(r)}(E)=\mathrm{Tr}[\Gamma_{l(r)} G^R \Gamma_{r(l)} G^A]$ and  $T^{\prime}_{l(r)}(E)=\mathrm{Tr}[\Gamma_{l(r)} G^R \Gamma_{c} G^A]$ are transmission functions.}

\begin{figure}[!th]
		\centering
		\includegraphics[width=\textwidth]{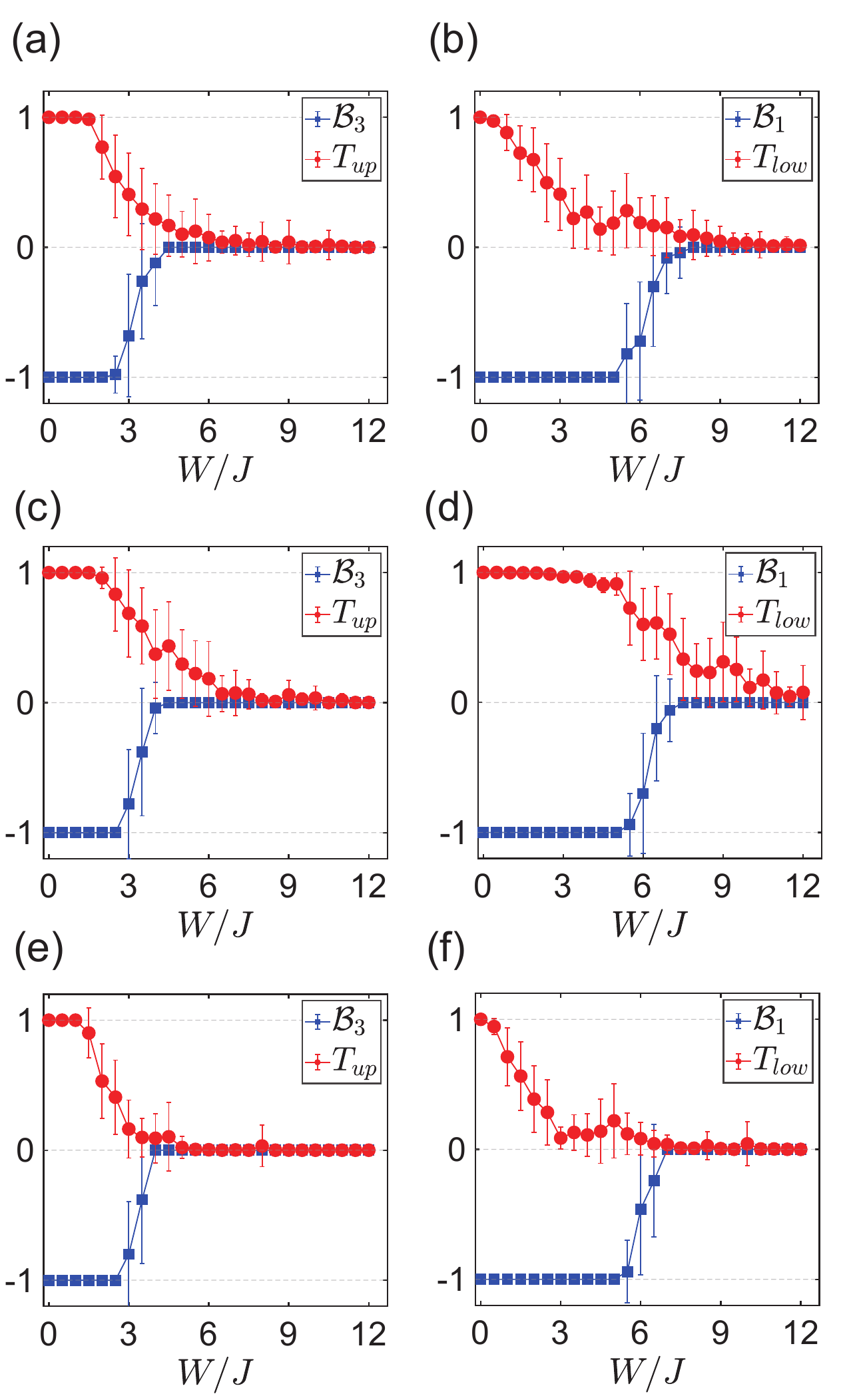}
		\caption{(a)(b) Magnon transmission $T$ calculated at the center of the (a) upper gap and (b) lower gap, compared with the Bott index of (a) the third band and (b) the first band. The size of (a) and (b) $20\times 20$. (c) and (d) $20\times 40$. (e) and (f) $40\times 20$.  Here $F=5J$, $K=20J$, and $\Delta=J^\prime=0$. The energies for transmission function calculation are $20.11J$ for lower gap and $29.65J$ for upper gap, respectively.}
		\label{kagome_size}
\end{figure}

\section{Size Effect of Topological Transition}\label{size_eff}
The effect of sample size on transmission can also be studied by the NEGF method. Figure \eqref{kagome_size} shows Bott index and the transmission across the whole sample for different sizes. When the width is small, the edge states at two edges may overlap each other, providing a back-scattering channel to diminish the transmission, so
that the transmission in the nontrivial phase becomes less than 1. With the increase of width, the overlap of edge states becomes smaller, and the drop of transmission occurs at a larger $W$, which can be seen by comparing Fig. \ref{kagome_size}(a)(b) and (c)(d). In other words, the localization becomes sharper at the nontrivial side. On the other hand, localized states may also contribute to the transmission when the sample is not long enough, giving finite transmission in the trivial phase. As the length increases, the localization becomes sharper at the trivial side, which is demonstrated by comparing Fig. \ref{kagome_size}(a)(b) and (e)(f).

\section{Details of Magnon Transport}\label{cur_detail}

\begin{figure}[!th]
		\centering
		\includegraphics[width=\textwidth]{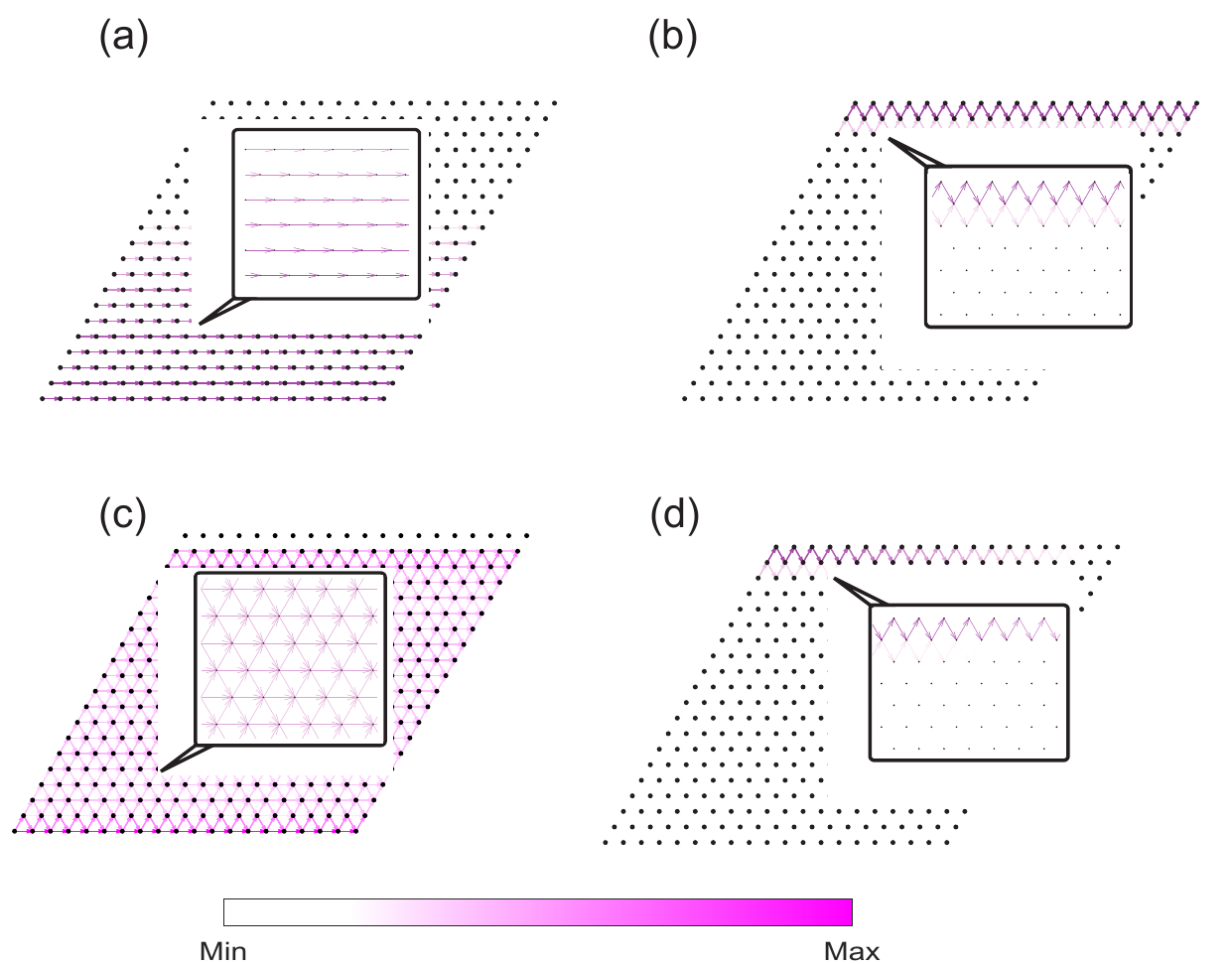}
		\caption{The illustration of local magnon currents for   \textcolor{black}{$J^{\prime}=\Delta=0$}, $F=5J$, and $K=20J$. Each site denotes a Kagome unit cell. The arrow direction denotes the direction of magnon currents, and the color of the arrows indicates the strength of magnon currents. (a) The lower topological edge modes with $E=19.86J$. (b) The upper topological edge modes with $E=29.65J$. (c) The bulk modes with $E=24.87J$. (d) The upper topological edge modes with $E=29.65J$ and $\alpha=2\times10^{-3}$.}
		\label{fig8}
\end{figure}
To observe magnon currents in detail, we calculate the distribution of magnon currents Eq. \eqref{Local current},
where $j_{mn}$ denotes the magnon current from $m$ site to $n$ site for specific $E$. $\eta_rH=\eta_rH_O$ is the real space Hamiltonian of OBC after Bogoliubov transformation. Figure \eqref{fig8} shows magnon currents of different modes with $D=J^{\prime}=0$. Each black dot denotes a Kagome unit cell and depth of the arrow color indicates the strength of magnon currents. Figure \eqref{fig8} (a), (b) and (c) show the local magnon currents lower, upper and bulk mode, respectively. The sample size is $N_x \times N_y =25 \times 25$. For $\alpha=0$, the magnitude of the magnon currents from left to right remains unchanged. The topological magnon current of the lower mode is localized at the lower edge, while the upper mode is localized at the upper edge, showing the opposite chirality among two edge states. When $\alpha\neq0$, magnon currents gradually dissipate in the sample during the transport process.

\begin{figure}[!th]
		\centering
		\includegraphics[width=\textwidth]{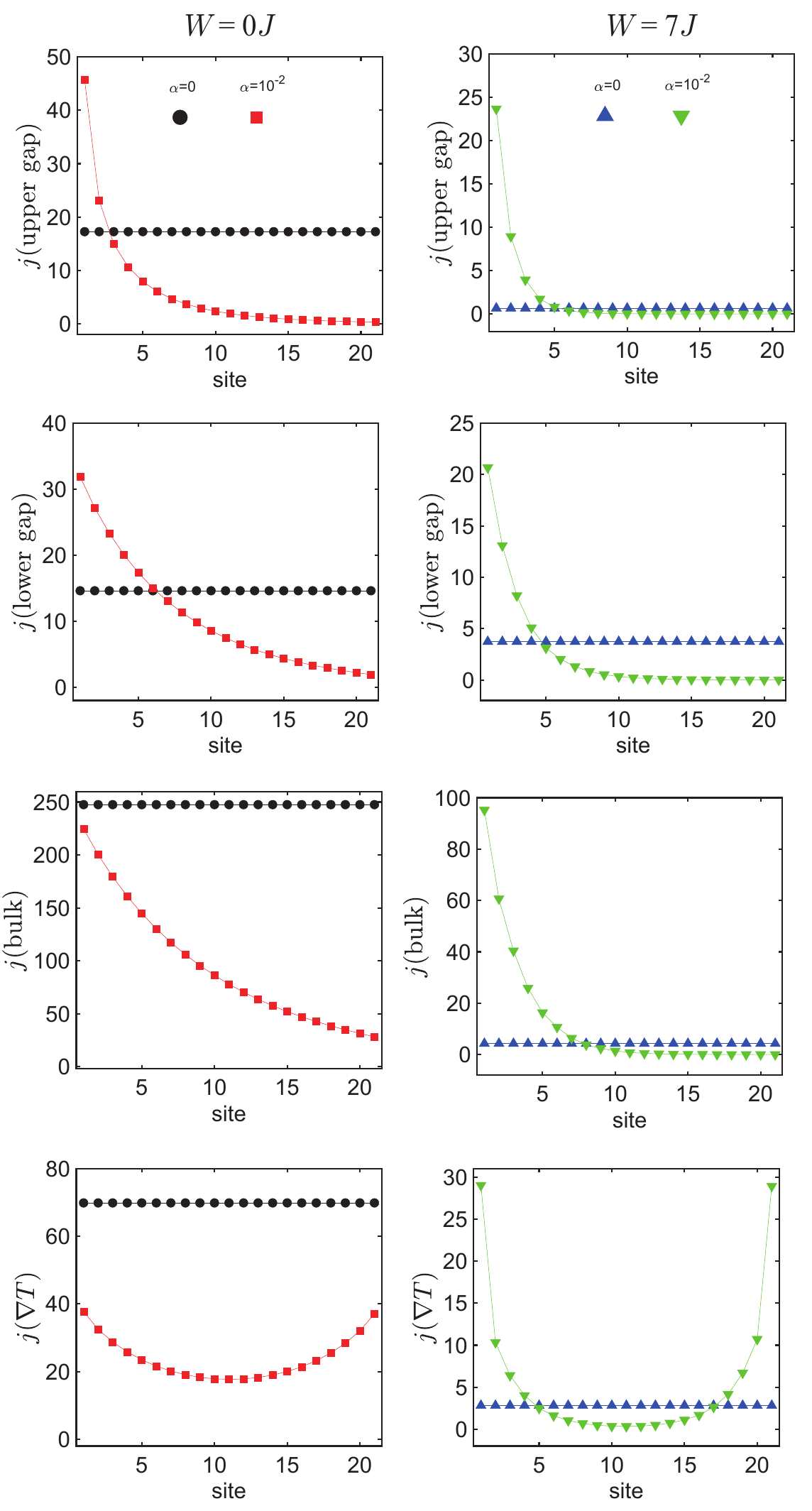}
		\caption{The magnon currents between each pair of adjacent columns for $W=0J$ (left) and $W=7J$ (right). The horizontal axis is the position ($n$ denotes the current from the $(n-1)$th column to $n$th column). (a) External coherent excitation at upper gap $E=29.65J$. (b) External coherent excitation at lower gap $E=20.11J$. (c) External coherent excitation in the bulk $E=14.5J$. (d) Temperature difference  \textcolor{black}{$t_l=1.2t_c$ and $t_r=0.8t_c$. Other parameters are $K=20J$, $F=5J$, and $J^{\prime}=\Delta=0$.}}
		\label{fig9}
\end{figure}

By replacing $N^B_L$ with $f$ in Eq.(\ref{lesser Green function}) , the local magnon currents from $m$ site to $n$ site  can also be calculated by Eq.\eqref{Local current} considering the external oscillating field excitation. To study the effects of disorder and damping, we calculate the magnon currents between each column. For 20 (long)$\times$80 (wide) samples, how the magnon currents depend on the position is shown in Fig.\eqref{fig9}. Site $n$ denotes the magnon currents from the $(n-1)$st column to the $n$st column [the left/right lead is treated as 0th/$(n+1)$th column], so that there are 21 data points from the left lead to the right lead. In the case of oscillating field excitation, the magnon currents remain uniform across all sites in the absence of damping.  When damping is present, the magnon currents decay progressively along the longitudinal direction, which is physically obvious. However, in the case of noncoherent thermal excitation, the magnitude of magnon currents decreases first and then increases, which is because the magnon current is mainly contributed by the temperature differences at the two sides. Despite this trend, $j_R$ is still slightly smaller than $j_L$ due to damping. In the present of disorder, the transport channels are destroyed and the magnitude of the magnon currents is reduced. The trend of how magnon currents depend on the propagating distance does not change qualitatively in spite of the presence of disorder. 

The value $\Lambda=22.5$ we use in the manuscript corresponds to a magnon generation rate $\int_0^{+\infty}L(E)dE\approx \int_{-\infty}^{+\infty}L(E)dE=\Lambda J=13.5\text{ meV}\approx 2\times 10^{13}\;\hbar$/s. This is a reasonable value and the reason is as follows. For $J=0.6$ meV from the experimental study \cite{Exp2021}, the magnon frequency in our study is in terahertz range (for example, the low gap at $E\sim20J$ corresponds to 2.9 THz). One of the most widely used THz source is the laser-pulse generated THz radiation in time-domain terahertz  spectroscopy (TDTS) \cite{THz}. Typical THz power is $\sim$1 $\mu$W and the pulse duration is $\sim$2 ps, so the total energy of a THz pulse is about 12.5 eV. However, the conversion efficiency from the driving electromagnetic wave to the excited spin wave is difficult to estimate. Here, $\sim0.1\%$ ($\approx$13.5 meV/12.5 eV) is a reasonable estimate, since in some theoretical studies \cite{transducer1,transducer2}, this efficiency can range from -40 dB ($\sim0.01\%$) to -3 dB ($\sim50\%$), depending on the design of the device. 

\bibliography{magnon.bib}

\end{document}